\documentstyle[12pt]{article} 

\begin{document}
 
\newcommand{\beq}{\begin{equation}}
\newcommand{\eeq}{\end{equation}}
\newcommand{\barr}{\begin{eqnarray}}
\newcommand{\earr}{\end{eqnarray}}

\newcommand{\andy}[1]{ }

\author{{\bf Fran\c{c}ois Goy}
\\
\\     
Dipartimento di Fisica,
Universit\^^ {a} di Bari \\
Via G. Amendola, 173 \\
I-70126  Bari, Italy \\ 
E-mail: goy@axpba1.ba.infn.it}

\title{{\bf Clock Synchronisation in the Vicinity of the Earth}}

\date{February 19, 1997}
 
\maketitle

\begin{abstract}
The transmission time of an electromagnetic signal in the vicinity of 
the earth is calculated to $c^{-2}$ and contains an orbital Sagnac term. 
On earth, the synchronisation of $TCB$ can be realised by atomic 
clocks, but not the one of $TCG$. The principle of equivalence is discussed. 
\end{abstract}
 

\setcounter{equation}{0}

\section{Introduction}
In 1994, Petit and Wolf \cite{pewo:94a} treated the 
problem of light 
propagation with the goal of ensuring picosecond accuracy for time 
transfer techniques using electromagnetic signals in the vicinity of the earth. 
They showed that the first post-Newtonian approximation of the geocentered 
metric, as defined by the 
resolution A4 of the International Astronomical Union (IAU) \cite{iau:92a}, 
was sufficient for their 
purpose and calculated the one-way time transfer of light to be applied when 
the spatial coordinates of the time transfer stations are known in a geocentric 
reference frame rotating with the earth. This expression was used to calculate 
at the same accuracy the special cases of the two-way and LASSO time transfers
via geostationary satellites. To the required accuracy, within a geocentric 
sphere of 200'000 km, the transmission delay is the sum of two terms: a 
geometrical one and a gravitational one. The geometrical one contains, under 
other, the Sagnac effect \cite{sagn:13a} due to the rotation of the earth on 
itself.

The aim of this article is to show that another effect arises in the 
one-way transmission time at the required accuracy: the 
Sagnac effect 
due to the orbital motion of the earth. In section 2, we 
present the reasons why such an effect takes place, the theoretical 
implications on the principle of equivalence as well as on the choice of 
a coordinate time scale on earth. In section 3, we calculate it 
explicitly in the same way as Petit and Wolf did in their article. In 
section 4, we discuss the implications on the realisability of the 
resolution A4 of the IAU.

\section{Orbital Sagnac effect}

The reason why Petit and Wolf did not take the orbital Sagnac effect into 
account is the following \cite{petit:96a}: `` I think (P. Wolf too) that what 
would be true in 
special relativity (i.e. if the reference frame bound to the earth would have 
for a reason X a displacement
similiar to the earth orbit and thus would be accelerated) is no more true in 
general 
relativity, where the earth is in free fall and only tidal effects
subsist, which we have evaluated and found $<$ 1 ps in our paper.''
We think that this point of 
view is in contradiction with some experiments for the reasons expounded in 
the following paragraphs. 

The Sagnac effect was observed 
around the earth using GPS satellites in common-view between 3 pairs of timing 
centers: Boulder, Tokyo and Braunschweig \cite{alla:85a}. Let us call $L$ the 
length covered by 
the light signal in this experiment. Saying that an around-the-world Sagnac 
effect occurs is just like saying that the time that light needs to do the 
round trip Boulder - Satellite 1 - Braunschweig -Satellite 2 - Tokyo 
- Satellite 3 - Boulder is not given by $L/c$ when measured in the rotating
(relative to distant radio sources) 
frame of the earth, but a Sagnac ``correction'' has 
to be applied and is equal in second order to $2\omega A_{E}/c^{2}$, where 
$A_{E}$ is the sum of the areas of the equatorial projections of the triangles
whose vertices are the center of the earth, the position of a timing center 
and a satellite and $\omega$ is the angular velocity of the earth.
$A_{E}$ is positive for signal propagation in the eastward 
direction and negative otherwise.

In order to prove that a Sagnac effect occurs also in free fall, we can 
imagine a slightly different experiment, where light signals are sent around 
the world between 3 satellites (the first one beeing geostationary), 
which are in  positions such that
the timing centers are on the straight line between two satellites.
Light makes the round trip: Satellite 1 - Timing center 1 -Satellite 2 - 
Timing center 2 -Satellite 3 - Timing center 3 - Satellite 1.  Just as in 
the case of the real experiment, a Sagnac effect will be measured also in this 
particular configuration, but in this case the timing centers play no role, 
since light could go directly from one satellite to the other, and we conclude 
that a Sagnac effect takes also place between the satellites only, in spite of 
the 
fact that they are in free fall. This conclusion is also valid if two 
satellites are so close to each other inside a region where it is possible to 
choose locally a freely falling inertial frame.
In post-Aristotelian physics, there is no 
reason to think, that there are different physical laws 
for satellites around the earth and the earth around the sun.

So, is that not true that on earth, because we are in free fall, we remain with 
tidal effects only? In our opinion, this is a widespread, but abusive 
interpretation of the principle of equivalence in general relativity. This 
last states in fact: `` At every space-time point in an arbitrary 
gravitational 
field it is possible to choose a ``locally inertial coordinate system'' such 
that, 
within a sufficently small {\em four-dimensional} region of the 
point in question, 
the laws of nature take the same form as in unaccelerated Cartesian 
coordinate systems in the absence of gravitation''.
The four-dimensionality of the neighbourhood is essential in this formulation 
\cite[p. 17]{ciwe:95a}, and 
refering to the earth as a region with negligible mass, we can only say 
that we are left with tidal effects only
in a sufficiently small three-dimensional region, during a sufficiently small 
time 
interval, but certainly not during a whole revolution of the earth around the 
sun.

Let us now consider the earth as a point of negligible mass rotating in 
circle around the 
sun and let us imagine an infinity of mirror-satellites on the earth orbit, 
maintaining a 
constant distance between them. These mirror-satellites would able us to 
send a light ray 
from earth, which would be constrained to follow the circular orbit and 
come back on earth. The reader familiar with the Sagnac effect, will recognize 
the simplified configuration used for pedagogical purposes. Using a barycentric
reference 
system ($BRS$) centered on the sun and non-rotating respective to the 
distant radio sources, we calculate easily, 
that 
a light ray sent on the orbit will come back after a time delay 
$\Delta t$ given by:
\andy{guevara}
\beq
\Delta t = \frac{L}{c} + \frac{2A_{S}\omega}{c^{2}}+O(c^{-3})
=\frac{L}{c} \pm \frac{L v_{E}}{c^{2}}+O(c^{-3})\;\;,
\label{eq:guevara}
\eeq
where $L$ is the length of the earth orbit, $A_{S}$ is the oriented area of 
the earth 
orbit, ${\bf \omega}$ the rotation vector of the earth around the sun and
$v_{E}$ is the velocity of the earth.
From (\ref{eq:guevara}), we find that
the velocity of light around its orbit is not $c$, 
but in first 
approximation $c\mp v_{E}$, when measured on earth.
Following the principle of 
equivalence, we can consider that in every point of the orbit at the moment 
of the 
passage of the light ray, there is a four-dimensional local coordinate system 
going at the 
velocity of the mirror. Then, in every of this frame the special relativity 
theory (SRT) is valid and in particular the velocity of light is $c$. This 
means that 
a light ray would take the time $\Delta t = L/c$ to make a round trip and 
negate the Sagnac effect. 

The problem above is nothing else than the transposition of the 
problem of the rotating platform from flat space-time to the curved 
space-time of general relativity. As 
already stressed by Selleri \cite{sell:96a}(see also \cite{gose:96a,fgoy:96a}), 
SRT applied locally on every small
tangential part of a rotating platform is unable to explain the Sagnac effect 
on this 
platform. Langevin considered also the problem long ago \cite{lang:21a}. He 
gave two different 
explainations of the Sagnac effect on the rotating platform. The first one is 
to 
consider a global central time such as the time of the laboratory. Time is then
defined everywhere and self-consistently but 
such a synchronisation of clocks gives a velocity of light on the rim of the 
platform 
which is not constant but in first approximation $c-\omega r$ and $c+\omega r$,
where $\omega$ is the angular velocity of the platform and $r$ its radius.
The second ``solution'' is to consider that the velocity 
of light is $c$ on the rim of the platform. This is equivalent to choose a 
local time, which is
given by integration of the differential of
the global time plus a differential contribution expressed in the 
space variables. The problem is that this contribution is not a total 
differential. Thus, the synchronisation procedure is path dependent and time 
cannot be defined consistently.
This problem has been recognized
on earth, by members of the timing community \cite{ashb:93a} :  
``This means that Einstein synchronization in a 
rotating reference frame 
is not self-consistent.[...]. In order to avoid difficulties with such 
non-transitivity it is best to adopt time in the {\em non-rotating} frame as 
the 
measure of time in the rotating frame.'' It can be easily proved that this 
problem occurs every time that the metric is non-static \cite{lali:59a}.

Despite the fact that the metric of the solar system is static, the 
problem 
is similar to the rotating platform. The global time in the solar system is
the Barycentric Coordinate Time  
($TCB$) \cite{iau:92a}. Let us consider two clocks at time $TCB$ located at 
${\bf x}$ and ${\bf x+dx}$ and
 rotating with velocity ${\bf v_{E}}$ on a circular orbit 
around the sun which are 
synchronised by adopting the time of the non-rotating frame. Then, the one-way 
velocity of light between the two clocks will be in first approximation 
$c\pm v_{E}$. If we use a local time, that is the Geocentric Coordinate Time
($TCG$), the velocity of light will
be $c$ in first approximation.
Comparing two $TCB$-synchronous events in points 
${\bf x}$ and ${\bf x} + {\bf dx}$, we see that they are not 
$TCG$-synchronous, but we have, using the transformation given in 
\cite{iau:92a}:
\andy{makno}
\barr  
TCB({\bf x})-TCG({\bf x})&=&\frac{1}{c^{2}}\int_{TCB_{0}}^{TCB}
 (v_{E}^{2}/2+U_{ext}({\bf x}))dt \nonumber \\
TCB({\bf x}+{\bf dx})-TCG({\bf x}+{\bf dx})&=&\frac{1}{c^{2}}
\left[\int_{TCB_{0}}^{TCB}
 (v_{E}^{2}/2+U_{ext}({\bf x}+{\bf dx}))dt \right. \nonumber \\
& &\left. + {\bf v_{E}}\cdot{\bf dx}\right] \;\;,
 \label{eq:makno}
\earr
so that:
\andy{chomsky}
\beq
\Delta TCG -\Delta TCB=\Delta TCG \cong -{\bf v_{E}}\cdot {\bf dx}/c^{2}\;\;,
\label{eq:chomsky}
\eeq
where the geocentric reference system
($GRS$) is centered on ${\bf x}$ for simplicity, $\Delta$ expresses the 
difference of time between the two points and 
 from (\ref{eq:makno})
to (\ref{eq:chomsky}) the difference of the external potential $U_{ext}$ 
between ${\bf x}$ and ${\bf x+dx}$ can be neglected.
Trying to extend $TCG$ spatially out of a local domain, for example 
by synchronising 
clocks along the orbit, we see that the procedure is path dependent, because 
${\bf v_{E}}\cdot {\bf dx}=\omega r^{2}d\theta$ is not a total differential 
in $r$ and $\theta$, where $r$ and $\theta$ are the polar coordinates of 
${\bf x}$ in $BRS$. 
If we now try to extend $TCG$ temporally, we can imagine 
two clocks, fixed in $GRS$, that is which have a constant direction respective 
to distant radio sources. Applying two times 
(\ref{eq:chomsky}) (that is to
$GRS$ at the point ${\bf x}(t)$ and later at point ${\bf x(t'>t)})$ to 
two pairs of $TCB$-simultaneous events, we see that the simultaneity 
in the two 
successive $GRS$ frames is not the 
same. So, $TCG$-synchronised clocks in the first $GRS$ frame would have to 
be resynchronised in order to 
obey to the different simultaneity of the second $GRS$. We will consider again 
this problem in section 4.
 So $TCG$ has highly undesiderable properties and it is 
better 
to consider, on the moving earth, that the simultaneity is given by the 
coordinate synchronisation of $TCB$.

This solution, which is the only one allowing a global and self-consistent 
definition of time, is also allowed by the principle of equivalence, 
which only states that it is {\em possible} to choose a local coordinate 
system such 
that the law of SRT are valid, but does not avoid to make an other choice
such that the velocity of light is not invariant.  More precisely: If in any 
point of a four dimensional surface, with general coordinates $x^{\alpha},\; 
(\alpha =0,1,2,3)$, 
there is a local transformation $x \rightarrow \xi$ such 
that the metric is 
locally Minkowskian, we can do a further resynchronisation of clocks $\xi^{0}
\rightarrow \xi'^{0}(\xi^{0},\xi^{i})$ and 
$\xi^{i}\rightarrow\xi'^{i}=\xi^{i},\;(i=1,2,3)$.
In particular it is possible to choose a transformation such that the time is 
globally defined 
on the whole surface, because $x^{0}$ is already global.
 If the coordinates of the velocity of the unprimed sytem 
are written in a ``vector'' form: ${\bf v}=\frac{d{\bf x}}{dt}$, this 
transformation is given by: 
$\xi^{0} \rightarrow \xi'^{0}=\xi^{0}+{\bf v}\cdot \mbox{\boldmath $\xi$}/c$, 
where 
$\mbox{\boldmath $\xi$}=(\xi^{1}, \xi^{2}, \xi^{3})$. 
 The laws of physics in the primed (local) system are 
the laws of the inertial theory \cite{sell:96a},
a theory equivalent to special relativity in inertial systems but maintaining 
an 
``absolute'' simultaneity \cite{mase:77a}. The  
metric expressed in the primed local coordinates is given by \cite{fgoy:97a}:
\andy{rosa}
\barr
g_{\alpha \beta}= \left(\begin{array}{cccc}
-1        & v_{1}/c             & v_{2}/c             &  v_{3}/c \\
v_{1}/c   & 1-v_{1}^{2}/c^{2}   & -v_{1}v_{2}/c^{2}   &  -v_{1}v_{3}/c^{2}\\
v_{2}/c   & -v_{1}v_{2}/c^{2}   & 1-v_{2}^{2}/c^{2}   &  -v_{2}v_{3}/c^{2}\\
v_{3}/c   & -v_{1}v_{3}/c^{2}   & -v_{2}v_{3}/c^{2}   &  1-v_{3}^{2}/c^{2}
\end{array}\right)
\label{eq:rosa}
\earr  
The spatial metric $\gamma_{ij}\;(i,j=1,2,3)$, which may be calculated from
$\gamma_{ij}=g_{ij}-g_{0i}g_{0j}/g_{00}$ is equal to $diag(1,1,1)$, as one 
can calculate from (\ref{eq:rosa}).

\section{One-way transmission time}

Let us now calculate the one-way transmission time of an electromagnetic signal
 between two points $a$ and $b$, given  
by their coordinates ${\bf x_{ra}}$ and ${\bf x_{rb}}$
in the geocentric rotating frame of the earth. Following Petit and 
Wolf, we calculate it in the non-rotating frame 
(coordinates: ${\bf x_{a}},\;{\bf x_{b}}$), where the path is a straigth 
line, but with the clock synchronisation of $TCB$, which allows a global 
definition of time. We suppose that all clocks are rate corrected for their
velocities and for the gravitational potential at their position so that they 
run at the rate of $TT$.
 One can show \cite{pewo:94a}, that up to 
an accuracy of order $O(c^{-3})$, the transmission time $T_{t}$, can be 
written as the sum of a geometrical $(T)$ and a gravitational term $(T_{g})$. 
That is:
$T_{t}=T+T_{g}$. $T_{g}$ is of order $c^{-3}$ and the choice of the coordinate
synchronisation of $TCB$ rather than the one of $TCG$, introduces corrections 
that are 
negligible (of order  $c^{-5})$, so that $T_{g}$ is given by (14) of 
\cite{pewo:94a}. Afterwards we will consider only the geometry and limit 
ourself to a calculation of order $c^{-2}$, since differences with 
\cite{pewo:94a} appear already at this order of approximation.
An electromagnetic signal is sent at
time $t_{0}$ from $a$ and arrives at time $t_{1}$ in $b$. From (\ref{eq:rosa}),
writing: $ds^{2}=g_{\alpha \beta}dx^{\alpha}dx^{\beta}=0$ and solving the 
resulting quadratic equation in $dx^{i}/dt\; (i=1,2,3)$, one obtains
the velocity of light in a direction
${\bf \hat{n}}$:
\andy{ravachol}
\beq
{\bf c}({\bf\hat{n}})=\frac{c{\bf\hat{n}}}
{1+\frac{{\bf v_{E}}\cdot {\bf \hat{n}}}{c}}=
{\bf\hat{n}}(c-{\bf v_{E}}\cdot {\bf \hat{n}}
+O(c^{-1}))\;\;,
\label{eq:ravachol}
\eeq
where ${\bf v_{E}}$ is the velocity of the earth in $BRS$.
The 
transmission time is given by:
\andy{cochon}
\beq
T=(1-U_{g}/c^{2})\mid {\bf x}_{rb}(t_{0})-{\bf x}_{ra}(t_{0})\mid/
c({\bf \hat{n}}) +s\;\;,
\label{eq:cochon}
\eeq
where $s$ represents the time taken by the signal to traverse the extrapath
due to the motion of $b$ in the non-rotating frame during the transmission, and
the factor $(1-U_{g}/c^{2})$ with $U_{g}/c^{2}=L_{g}$ arises because 
$T$ is measured in units of $TT$ (see \cite{iau:92a}:
$L_{g}= 6.969291\cdot10^{-10}$).
Defining:
\andy{kropotkine}
\barr
{\bf R_{0}}&=&{\bf x_{rb}}(t_{0})-{\bf x_{ra}}(t_{0}) \nonumber \\
{\bf v_{b}}&=&\mbox{\boldmath $\omega$}\times {\bf x_{rb}} + {\bf v_{rb}} 
+O(c^{-2}) \;\;,
\label{eq:kropotkine}
\earr
with ${\bf v_{rb}}$ being the velocity of b 
in the rotating frame and the two frames having their axes pointing in the same 
direction at $t=t_{0}$.

The path travelled by the signal ${\bf R}(T)$ can be expressed as a series 
expansion in terms of $T$ in the non-rotating frame:
\andy{lamort}
\beq
{\bf R}(T)= {\bf R_{0}}+{\bf v_{b}}T+O(T^{2})
\label{eq:lamort}
\eeq
Calculating its magnitude $R(T)$, 
expanding the square root and writing  $R(T)$$=$
$c({\bf \hat{n}})T/(1-U_{g}/c^{2})$
we find:
\andy{bonot}
\beq
T=[R_{0}+({\bf R_{0}}\cdot{\bf v_{b}}/R_{0})T
+O(T^{2})](1-U_{g}/c^{2})/c({\bf \hat{n}})
\label{eq:bonot}
\eeq
Starting with $T=R_{0}/c({\bf \hat{n}})$ and iterating once yields an 
expression for the transmission time in terms of the known quantities
${\bf R_{0}}$ and ${\bf v_{b}}$:
\andy{bader}
\beq
T = R_{0}/c+ {\bf R_{0}}\cdot({\bf v_{E}}+{\bf v_{b}})
/c^{2} +O(c^{-3})
\label{eq:bader}
\eeq
and Petit and Wolf found at second order in $c$:
\andy{poubelle}
\beq
T^{TCG} = R_{0}/c+ {\bf R_{0}}\cdot{\bf v_{b}}
/c^{2} +O(c^{-3})
\label{eq:poubelle}
\eeq
Strictly speaking, the transmission time calculated in (\ref{eq:bader}) does 
not 
correspond to the same physical situation as the transmission time calculated 
in (11)
The reason is the following: ${\bf R_{0}}$ 
is not the same in both situations since it depends on the choosen 
synchronisation. In theory, the length between to points is measured with 
unit rods between the simultaneous positions of these two points. If these 
points 
are fixed in a given reference frame the notion of simultaneity does not 
influence the result of a measure in this frame \cite{anst:94a}. 
But if the points are moving, 
this is no more true because the same numerical value of the 
coordinate time in different synchronisations does not correspond to the same 
position of the moving objects and inversely the points are at the same places 
at different times in different synchronisations. 
From $t^{TCB}\cong t^{TCG}+{\bf v_{E}}\cdot{\bf x}/c^{2}$ and 
(\ref{eq:kropotkine}), we calculate easily:
\andy{aristide}
\barr
{\bf R_{0}^{TCB}}&={\bf R_{0}^{TCG}}& + c^{-2}\left[{\bf v_{rb}^{TCG}}
   ({\bf v_{E}}\cdot {\bf x_{rb}^{TCG}}(t_{0}))\right. \nonumber \\
  & & -\left. {\bf v_{ra}^{TCG}}
   ({\bf v_{E}}\cdot {\bf x_{ra}^{TCG}}(t_{0}))\right] + O(c^{-4})
\label{eq:aristide}
\earr
The velocities are also not the same and vary at second order.
\andy{ntm}
\beq
{\bf v_{b}^{TCB}} = {\bf v_{b}^{TCG}} (1 -
{\bf v_{E}}\cdot {\bf v_{b}^{TCG}}/c^{2})
 + O(c^{-4})
\label{eq:ntm}
\eeq
Substituing ${\bf R_{0}}$ and ${\bf v_{b}}$ of (\ref{eq:bader}) by their 
value of
(\ref{eq:aristide}) and (\ref{eq:ntm}) does not change the result at this order 
of approximation. So, we can say that (\ref{eq:bader}) and (\ref{eq:poubelle}) 
are calculated in the same physical situation, at $O(c^{-3})$,
but with different synchronisations.
The additional term in 
(\ref{eq:bader}) respective to (\ref{eq:poubelle}) comes from 
(\ref{eq:ravachol}) and is the orbital Sagnac effect. It is given in first 
approximation by
${\bf R_{0}}\cdot{\bf v_{E}}/c^{2}$ and does not occur if ${\bf R_{0}}$ is 
perpendicular to the orbital velocity of the earth. If ${\bf R_{0}}$ is 
parallel 
to ${\bf v_{E}}$ it amounts 333 ns for a link of 1000 km. Of 
course the presence this new term respective to the expression of Petit and 
Wolf is only measurable if we adopt the synchronisation of $TCB$.
The question is then: will such an orbital 
effect arise if we do not adopt this ``convention''? 

\section{Is the syn\-chro\-ni\-sa\-tion of TCG re\-alis\-able on earth}

If we use the coordinate synchronisation of $TCG$ on earth as recommended by 
the IAU, the orbital Sagnac effect will not be measured 
immediatly, but a desynchronisation of clocks will occur. 
This is due to the fact that two clocks $A$ and $B$, located in ${\bf x_{A}}$ 
and ${\bf x_{B}}$, which are Einstein 
synchronised in an inertial frame do not remain Einstein-synchronous after or 
during acceleration \cite{mast:93a}. For the same reasons, that are expounded 
in section 2, this fact is not only true within the frame of special relativity
but also for two freely falling clocks, close to each other and in orbit 
around the sun.
The desynchronisation of clocks is given 
by \cite{ashb:93a}:
\andy{genet}
\beq
\Delta t =t_{B}-t_{A}= -\frac{{\bf u}\cdot {\bf R}}{c^{2}}\;\;,
\label{eq:genet}
\eeq
where ${\bf u}$ is the relative velocity of the two frames and ${\bf R}=
{\bf x_{B}}-{\bf x_{A}}$ is the 
vector separating the two clocks. So if realise $TCG$ at 
time $t_{0}=0$ and we call the corresponding non-rotating frame $GRS_{0}$, 
then the velocity 
${\bf u}$ of the earth relative to $GRS_{0}$ will be for a circular orbit:
\andy{meinhof}
\beq
{\bf u}=v_{E}\left(\begin{array}{c}-\sin\theta \\ \cos\theta-1\end{array}\right)
\;\;,
\label{eq:meinhof}
\eeq
where $v_{E}$= 30 km/s is the velocity of the earth relative to the sun, and 
$\theta$ is the angle between the two positions of the earth with vortex on the 
sun. If the two clocks are making an angle $\phi$ with the x-axis at time
$t$, then (\ref{eq:genet}) and 
(\ref{eq:meinhof}) give:
\andy{fasel}
\beq
\Delta t = \frac{v_{E}R}{c^{2}}[\sin(\phi -\theta)-sin(\phi)]\;,
\label{eq:fasel}
\eeq
the x-axis being given by the straight line between the sun and the earth at 
time $t_{0}$. 

For two clocks, that are fixed on earth $\phi$ has in general a periodic 
diurnal variation and $\theta$ an annual period. In this particular case, which 
could be tested, we obtain from (\ref{eq:fasel}) a diurnal variation,
modulated in amplitude by an annual one. The maximum of the desynchronisation
will be attained after 6 months, where the one-way velocity of light measured 
between the two clocks can change in 12 hours from $c+v_{E}$ to $c-v_{E}$, if 
the link is parallel to the earth velocity. This effect 
can be distinguished from temperature delays \cite{imki:92a}
which share the same period, but 
can be dephased and do not have the same caracteristics. For a link of 1000km,
$\Delta t$ amount 
666 ns after 6 months if the link 
is parallel to 
the earth velocity. In order to detect it, we need a link and  clocks with
a relative precision  of $4\cdot10^{-14}$ over 6 months.

The resolution A4 of the IAU contains the description of various time scales 
and coordinate systems and for this reason almost already all elements of a 
correct theoretical description of time on earth, such as realised by a net of 
atomic clocks. The synchronisation of $TCG$ or $TT$ cannot be
realised by atomic clocks working continuously without resynchronisation, but 
the synchronisation of $TCB$ can be realised, because $BRS$ can be 
considered to be not accelerated respective to the universe. Thus, 
the metric of a geocentered non-rotating reference frame
is non-static and non-stationary, already at order $c^{-1}$, because of 
the presence of the terms $g_{0i}={v_{E}}_{i}/c$, which are varying in time. 
In principle, we should also consider
the motion of the whole solar system around the galaxy, and following our 
reasoning $TCB$-synchronised clocks will slowly desynchronise with a maximum 
of 5 $\mu s$ for a link of 1000 km after 250 millions of years, but it can be 
considered as negligible.

\section{Conclusion} 

It was shown that the natural behaviour of atomic clocks in the vicinity of the 
earth is not described correctly by the synchronisation of $TCG$. Aside from 
gravitational effects of the earth, $GRS$ corresponds to 
the application of the principle of equivalence to the earth in free fall 
around the sun. But in the principle of equivalence the four-dimensionality 
is important and thus $TCG$ 
can only be realised during a sufficiently small time interval in a 
three-dimensional neighbourhood of the earth. An extension of $TCG$ out of this 
neighbourhood leads to a non-transitive and not self-consistent clock 
synchronisation around the earth orbit and to an annual and diurnal 
desynchronisation of clocks on earth. The author proposes that we adopt the 
synchronisation of $TCB$, the coordinate time of the solar system as time 
on earth, which leads to a correct description of the behaviour of clocks on 
earth. The principle of equivalence is reinterpreted in the light of the works 
on the conventionality of clock synchronisation in inertial frames:
A global and self-consistent definition of time 
implies that the laws of 
physics in a locally (four-dimensional) freely falling system are those of 
the inertial theory and in 
particular the one-way velocity of light is not invariant on earth. 
In a sense, the work done here is nothing 
more than the extension to the solar system of what had already been done by 
physicists of the timing community: the synchronisation in the rotating frame 
of the earth is given by the synchronisation in the non-rotating frame.
Note finally, that the author does not consider the work done here as 
definitive. The increasing precision of clocks makes necessary to calculate the 
one-way transmission time at the order $c^{-3}$.

\section{Acknowledgements}
I want to thanks the Physics Departement of Bari Uni\-ver\-si\-ty for 
hospitality, 
Prof. F. Selleri for its kind suggestions and criticisms, the Swiss 
National Science Foundation and the Swiss Academy of Engineering Sciences 
for financial support and 
the sun for inspiration.

\end{document}